\documentclass[11pt]{article}
\usepackage{epsfig}
\begin{document}
\begin{center}
\vskip 1.0cm

\bigskip {\Large Stochastic Analysis of Subcritical Amplification of
Magnetic Energy }

{\Large in a Turbulent Dynamo}

\vskip 1.0cm

Sergei Fedotov$^{1,3}$ , Irina Bashkirtseva$^{2}$ and Lev Ryashko$^{2}$

$^1$ Department of Mathematics, UMIST - University of Manchester

Institute of Science and Technology, Manchester, M60 1QD UK,

$^2$ Department of Mathematical Physics, Ural State University,

Lenin Av., 51, 620083 Ekaterinburg, Russia.

\

$^3$ Author to whom correspondence should be addressed.

\end{center}

\vskip 0.5cm

\begin{center}
\bigskip Abstract
\end{center}

We present and analyze a simplified stochastic $\alpha \Omega -$dynamo model
which is designed to assess the influence of additive and multiplicative
noises, non-normality of dynamo equation, and nonlinearity of the $\alpha -$%
effect and turbulent diffusivity, on the generation of a large-scale
magnetic field in the subcritical case. Our model incorporates random
fluctuations in the $\alpha -$parameter and additive noise arising from the
small-scale fluctuations of magnetic and turbulent velocity fields. We show
that the noise effects along with non-normality can lead to the stochastic
amplification of the magnetic field even in the subcritical case. The
criteria for the stochastic instability during the early kinematic stage are
established and the critical value for the intensity of multiplicative noise
due to $\alpha -$fluctuations is found. We obtain numerical solutions of
non-linear stochastic differential equations and find the series of phase
transitions induced by random fluctuations in the $\alpha -$parameter.

\newpage

\section{Introduction}

The understanding of the generation and maintenance of a large
scale magnetic field in astrophysical objects is a problem of
exceptional importance and difficulty. It is widely accepted that
the magnetic field is generated by the turbulent flow of the
electrically-conducting fluid. Inhomogeneous velocity fluctuations
stretch magnetic lines and amplify the magnetic field. These small
scale fluctuations of turbulent flow are primarily responsible for
the generation of magnetic fields. The problem is that it is
difficult to resolve them. Thus their influence on the resolved
large-scale magnetic field has to be modelled. A traditional
closure scheme is based on the $\alpha -$effect according to which
small-scale fluctuations can be described by an average term
involving the curl of the
mean magnetic field, $\mathbf{B,}$ written as $\nabla \times (\alpha \mathbf{%
B)}.$ The asymptotic analysis of the induction equation exploiting the
assumption of two separated scales for turbulent flow leads to the effective
macroscopic equation for the large scale magnetic field $\mathbf{B}(t,%
\mathbf{x})$ [1-5]
\begin{equation}
\frac{\partial \mathbf{B}}{\partial t}=\nabla \times (\alpha \mathbf{B}%
)-\nabla \times (\beta \nabla \times \mathbf{B)}+\nabla \times (\mathbf{u}%
\times \mathbf{B}),  \label{dynamo}
\end{equation}
where $\ \mathbf{u}\ $ is the mean velocity field of the turbulent
flow, $\ \alpha \ $ is the coefficient of the $\alpha $-effect and
$\beta $ is the turbulent magnetic diffusivity. Traditionally the
phenomenon of generation of magnetic fields was analyzed by
considering perturbations of the trivial state $\mathbf{B}=0$ and
looking for exponential solutions to the deterministic PDE
(\ref{dynamo}) with appropriate boundary conditions. While this
standard stability analysis successfully predicts the dynamo
action for the supercritical case, there are situations for which
this eigenvalue analysis fails to predict the subcritical onset of
instability [6-10]. It was pointed out in \cite{FI1} that the
closure scheme involving only deterministic $\alpha \beta
-$parameterization is not completely satisfactory since unresolved
fluctuations may produce random terms on the right hand side of
the dynamo equation (\ref{dynamo}). Moreover, it follows from
astronomical observations that large scale magnetic fields exhibit
a rich random variability both in space and time that cannot be
described by the deterministic equation (\ref {dynamo}).

The importance of noise effects in the dynamo problem has been recognized
previously and several attempts have been made to account for the effects of
spatial and temporal fluctuations in small scale magnetic and velocity
fields on the generation process. Kraichnan considered fluctuations in the $%
\alpha $-parameter and found a negative contribution \ to turbulent
diffusivity from helicity fluctuations \cite{Kr}. Hoyng with colleagues in
[12-14] studied in detail the effect of random fluctuations in the $\alpha $%
-parameter by considering the system of stochastic linear equations for
eigenmodes corresponding to the dynamo equations. They found the excitation
of those modes such that their magnetic energy is proportional to $\gamma
^{-1},$ where $\gamma $ is the damping rate. Stochastic dynamics of magnetic
field generation have been also analyzed by Farrell and Ioannou in \cite{FI1}%
, where they examine a mechanism by which small-scale fluctuations
excite the large scale magnetic field. They modelled these
fluctuations by an additive noise term in the mean field equation
and identified the crucial role of non-normality on the dynamo
process. Numerical simulations of the magnetoconvection equations
were performed in \cite{Proctor} with analysis of effects of noise
and non-normal transient growth. Inhomogeneous turbulent helicity
fluctuations were considered in \cite{Sil}. One should mention the
dynamo model that exhibits aperiodic switching between regular
behavior and chaotic oscillation \cite{PST2}. Stochastic dynamo
theory, using the term of an incoherent dynamo, was proposed by
Vishniac and Brandenburg in \cite{VB}. They showed how random
fluctuations in the
helicity can generate a large-scale magnetic field for the $\alpha \Omega -$%
dynamo. We note that this model is closely related to the present
work. However, they did not consider the transient amplification
due to the non-normality of the dynamo equation. It turns out that
non-normal dynamical systems exhibit \ an extraordinary
sensitivity to random perturbations which leads to great
amplification of the second moments of the stochastic dynamical
systems (see \cite{Farrel1,Grossmann,Fedotov}). Our recent work
has demonstrated the possibility for stochastic magnetic energy
amplification in the subcritical situation where the dynamo number
is less than critical \cite {F1}. These observations motivate
further studies of noise and nonlinearity
effects which we consider below in the context of a simplified stochastic no-%
$z$ model (see \cite{Moss}). An issue we address in this paper is how the
random fluctuations may be appropriately incorporated into the classical $%
\alpha \Omega -$dynamo model. Our analysis especially focuses upon the
multiplicative noise due to random fluctuations in the $\alpha $-parameter
and the non-normality of the dynamo equation operator. Recall that an
operator is said to be non-normal if it does not commute with its adjoint in
the corresponding scalar product. The determination of the effect of the
noise, along with non-normality, on the amplification of magnetic energy
during the early kinematic stage is the primary contribution of this paper.

In section II, we discuss the deterministic $\alpha \Omega -$dynamo model,\
its equilibrium points and transient growth effects. In section III, we
consider a linear stochastic model for the subcritical case. We derive
equations for the second moments and find their stationary values. We
demonstrate the important differences between the non-normal system and the
normal one under the influence of additive noise. We explore the influence
of multiplicative noise due to fluctuations in the $\alpha $-parameter on
the amplification of magnetic energy in the subcritical case during the
kinematic stage. We derive the criteria under which the second moments grow
exponentially with time (kinematic regime). Finally, in section IV, we
perform numerical experiments showing that there exists a series of
noise-induced phase transitions in the $\alpha \Omega -$dynamo model.

\section{Stochastic $\protect\alpha \Omega -$dynamo model in a thin-disk
approximation}

Following \cite{ZRS,RShS} we consider here the thin-disk approximation to
the dynamo equation for spiral galaxies, in which a turbulent disk of
conducting fluid of uniform thickness $\ 2h\ $and radius $\ R\ $($R\gg h$)
rotates with angular velocity $\ \Omega (r)$. We restrict ourselves to the
case of the$\ \alpha \Omega -$dynamo for which the differential rotation
dominates over the $\alpha $-effect. The governing equations for the
components of the axisymmetric magnetic field, $\ B_{r}\left( t,r,z\right) \
$and $\ B_{\varphi }\left( t,r,z\right) \ $ in the polar cylindrical
coordinates $\left( r,\varphi ,z\right) $\ can be written as,
\[
\frac{\partial B_{r}}{\partial t}=-\frac{\partial }{\partial z}(\alpha
(t,z,r)B_{\varphi })\ +\beta \frac{\partial ^{2}B_{r}}{\partial z^{2}}+\frac{%
\beta }{r}\frac{\partial }{\partial r}\left( \frac{1}{r}\frac{\partial B_{r}%
}{\partial r}\right) +\ f_{r}(t,z,r),
\]
\begin{equation}
\frac{\partial B_{\varphi }}{\partial t}=g_{\omega }\ B_{r}+\beta \frac{%
\partial ^{2}B_{\varphi }}{\partial z^{2}}+\frac{\beta }{r}\frac{\partial }{%
\partial r}\left( \frac{1}{r}\frac{\partial B_{r}}{\partial r}\right) +\
f_{\varphi }(t,z,r),\   \label{main}
\end{equation}
where $\ g_{\omega }=rd\Omega /dr\ $is the measure of differential rotation
(usually $rd\Omega /dr<0),$ and $\ f_{r}(t,z,r)$ and $\ f_{\varphi }(t,z,r)\
$are the stochastic terms describing unresolved turbulent fluctuations. The
components $B_{r}\left( t,r,z\right) \ $ and $\ B_{\varphi }\left(
t,r,z\right) $ obey vacuum boundary conditions on the thin disc surfaces $%
B_{r,\varphi }\left( t,r,-h\right) =0,\ B_{r,\varphi }\left( t,r,h\right) =0$
(see \cite{ZRS, RShS}).

In what follows we neglect the spatial structure of the magnetic field along
the radius $r$ and the height of the galaxy adopting the well-known no-$z$
model \cite{Moss}. The aim here is to concentrate on the studies of the
influence of random fluctuations and non-normality on the dynamo process. We
also take into account the non-linear backreaction \cite{b2,b3}. The
dynamical system for the azimuthal, $B_{\varphi }\left( t\right) ,\ $and
radial,$\ B_{r}\left( t\right) ,$ components of the magnetic field $\mathbf{B%
}$ can be written then in the form
\[
\frac{dB_{r}}{dt}=-\frac{\alpha (|\mathbf{B}|)(1+\xi _{\alpha }(t))}{h}%
B_{\varphi }-\frac{\pi ^{2}\beta (|\mathbf{B}|)}{4h^{2}}B_{r}+\xi _{r}(t),
\]
\begin{equation}
\frac{dB_{\varphi }}{dt}=g_{\omega }B_{r}-\frac{\pi ^{2}\beta (|\mathbf{B}|)%
}{4h^{2}}B_{\varphi }\,.  \label{governing}
\end{equation}
This dynamical system may be regarded as an one-mode approximation of (\ref
{main}). It should be noted that the deterministic critical conditions for
the generation of a magnetic field are the same for both models (\ref{main})
and (\ref{governing}): $\alpha h^{3}\beta ^{-2}|g_{\omega }|>\pi ^{4}/16$
\cite{RShS}. Here we introduce the non-linear functions $\alpha (|\mathbf{B}%
|)$ and $\beta (|\mathbf{B}|)$ describing the quenching of the $\alpha -$%
effect and turbulent magnetic diffusivity. It is well-known that the current
theories disagree about how the dynamo coefficients $\alpha (|\mathbf{B}|)$
and $\beta (|\mathbf{B}|)$ are suppressed by the mean field $\mathbf{B}$
\cite{b2,b3,b4,b5,b6}. Here we describe the dynamo quenching by using the
standard forms \cite{Widrow}
\begin{equation}
\alpha (|\mathbf{B}|)=\alpha _{0}\left( 1+k_{\alpha }(B_{\varphi
}/B_{eq})^{2}\right) ^{-1},\;\;\;\beta (|\mathbf{B}|)=\beta _{0}\left( 1+%
\frac{k_{\beta }}{1+(B_{eq}/B_{\varphi })^{2}}\right) ^{-1},
\label{nonlinear}
\end{equation}
where $k_{\alpha }$ and $k_{\beta }$ are constants of order one, and $B_{eq}$
is the equipartition strength. We note that for the $\alpha \Omega -$dynamo
the azimuthal component $B_{\varphi }\left( t\right) $ is much larger$\ $%
than the radial field$\ B_{r}\left( t\right) ,$ therefore, $\mathbf{B}%
^{2}\simeq B_{\varphi }^{2}.$ The issue of the strong dependence of $\alpha $
and $\beta $ on the magnetic Reynolds number $R_{m}$ is still controversial
and we did not include it in our analysis.

Regarding random perturbations in the system (\ref{governing}), we adopt a
phenomenological mesoscopic approach in which the parameters of dynamo
equations are assumed to be random functions of time \cite{Gardiner,LH}. The
multiplicative noise $\xi _{\alpha }(t)$ in (\ref{governing}) stands for the
rapid random fluctuations in the parameter $\alpha .$ One can show that they
are more important than the random fluctuations in the turbulent magnetic
diffusivity $\beta $ \cite{Hoyng,H2,H3}. The additive noise term $\xi
_{r}(t) $ represents the stochastic forcing arising from the small-scale
fluctuations of magnetic and turbulent velocity fields \cite{FI1}. Both
random terms are assumed to be independent Gaussian white noises with zero
means $<\xi _{\alpha }(t)>=0,$ $<\xi _{r}(t)>=0$ and correlations:
\begin{equation}
<\xi _{\alpha }(t)\xi _{\alpha }(s)>=2D_{\alpha }\delta (t-s),\;\;\;<\xi
_{r}(t)\xi _{r}(s)>=2D_{r}\delta (t-s),  \label{noise}
\end{equation}
where $D_{\alpha }$ and $D_{r}$ are the intensities of the noises and the
angular brackets , $<\cdot >,$ denote the statistical average. One can show
that the additive noise in the second equation in (\ref{governing}) is less
important and can be ignored. In particular, the main contribution to the
second moment of $B_{\varphi }$ comes from the additive noise in the first
equation (see, for example, \cite{Fedotov}). Of course, this paper addresses
the over-simplified case of magnetic field generation in galaxies.
Nonetheless, we present this work as an illustration of the influence the
random fluctuations and non-normality may play in the dynamo process, and
which therefore should be accounted for in more complicated dynamo modelling
like the stochastic PDE (\ref{main}).

The dynamical system (\ref{governing}) is well studied for the case when the
noise terms are absent, and there has been much progress in the prediction
of the growth rates induced by both the $\alpha -$effect and differential
rotation and the corresponding speed of magnetic waves \cite{MSS}. However,
there is considerably less understanding of generation of a magnetic field
in the presence of random fluctuations. Although the equations (\ref
{governing}) are a theoretical simplification of what really happens in
galaxies, we strongly believe that it does provide a useful framework for
understanding the interaction of stochastic perturbation, non-normality and
non-linear effects.

It is convenient to rewrite the governing equations (\ref{governing}) in a
nondimensional form by using an equipartition field strength $B_{eq},$ a
length $h$, and a time $\Omega _{0}^{-1},$ where $\Omega _{0}$ is the
typical value of angular velocity. In terms of the dimensionless parameters
\begin{equation}
g=\frac{|g_{\omega }|}{\Omega _{0}},\;\;\;\delta =\frac{R_{\alpha }}{%
R_{\omega }},\;\;\;\varepsilon =\frac{\pi ^{2}}{4R_{\omega }}%
,\;\;\;R_{\alpha }=\frac{\alpha _{0}h}{\beta },\;\;\;R_{\omega }=\frac{%
\Omega _{0}h^{2}}{\beta },  \label{parameters}
\end{equation}
the stochastic dynamo equations (\ref{governing}) can be written in the form
of SDE \cite{Gardiner,LH}
\begin{equation}
dB_{r}=-(\delta \varphi _{\alpha }(B_{\varphi })B_{\varphi }+\varepsilon
\varphi _{\beta }(B_{\varphi })B_{r})dt-\sqrt{2\sigma _{1}}\varphi _{\alpha
}(B_{\varphi })B_{\varphi }dW_{1}+\sqrt{2\sigma _{2}}dW_{2},  \label{basic1}
\end{equation}
\begin{equation}
dB_{\varphi }=-(gB_{r}+\varepsilon \varphi _{\beta }(B_{\varphi })B_{\varphi
})dt\ ,  \label{basic}
\end{equation}
where $W_{1}$ and $W_{2}$ are independent standard Wiener processes, and $%
\sigma _{1}$ and $\sigma _{2}$ are the noise intensities
\begin{equation}
\sigma _{1}=\frac{D_{\alpha }}{h^{2}\Omega _{0}},\;\;\;\sigma _{2}=\frac{%
D_{r}}{B_{eq}^{2}\Omega _{0}}.  \label{intensity}
\end{equation}
Here we introduce the functions $\varphi _{\alpha }(B_{\varphi })$ and $%
\varphi _{\beta }(B_{\varphi })$ describing non-linear quenching
\begin{equation}
\varphi _{\alpha }(B_{\varphi })=\frac{1}{1+k_{\alpha }B_{\varphi }^{2}}%
,\;\;\;\varphi _{\beta }(B_{\varphi })=\frac{1+B_{\varphi }^{2}}{1+(k_{\beta
}+1)B_{\varphi }^{2}}.  \label{saturation}
\end{equation}

Now let us discuss the other parameters in (\ref{basic1}),(\ref{basic}),
namely, $\delta =$ $R_{\alpha }/R_{\omega }$ and $\varepsilon =\pi
^{2}/4R_{\omega }.$ The parameter $\delta $ is the characteristic of
relative importance of the $\omega -$effect and the $\alpha -$effect. For
the $\alpha \Omega -$dynamo the differential rotation dominates over the $%
\alpha $-effect, that is, $R_{\alpha }\ll R_{\omega }.$ Moreover, the
diffusion time $h^{2}/\beta $ is much larger than the time $\Omega
_{0}^{-1}, $ therefore, both parameters $\delta $ and $\varepsilon $ are
small. Their typical values for spiral galaxies are $0.01-0.1$ ($R_{\omega
}=10-100,$ $\ R_{\alpha }=0.1-1)$ \cite{RShS}. It turns out that for small
values of $\delta $ and $\varepsilon $ , the linear operator (matrix) in (%
\ref{basic1}),(\ref{basic}) is a highly non-normal one, since $g\sim 1.$
This can lead to a large transient growth of the azimuthal component $%
B_{\varphi }\left( t\right) $ in the \textit{subcritical }case\textit{.} The
nonlinear interactions may lead to a further amplification of this small
disturbance \cite{F1}. The crucial idea behind subcritical transition is
that the $\alpha -$effect or the $\omega -$effect might be relatively weak,
but the generation and maintenance of the large scale magnetic field is
still possible.

\section{Deterministic system}

Let us briefly review the dynamics of the system (\ref{basic1}),(\ref{basic}%
) in the absence of noise terms. It takes the form
\[
\frac{dB_{r}}{dt}=-\delta \varphi _{\alpha }(B_{\varphi })B_{\varphi
}-\varepsilon \varphi _{\beta }(B_{\varphi })B_{r},
\]
\begin{equation}
\frac{dB_{\varphi }}{dt}=-gB_{r}-\varepsilon \varphi _{\beta }(B_{\varphi
})B_{\varphi }.  \label{deterministic}
\end{equation}
The equilibrium points of the system (\ref{deterministic}) can be found from
$\delta \varphi _{\alpha }(B_{\varphi })B_{\varphi }+\varepsilon \varphi
_{\beta }(B_{\varphi })B_{r}=0,$ $gB_{r}+\varepsilon \varphi _{\beta
}(B_{\varphi })B_{\varphi }=0.$ The stationary value of the radial component,%
$\ B_{r},$ can be expressed in terms of the azimuthal one, $B_{\varphi },$
\begin{equation}
B_{r}=-\frac{\varepsilon \varphi _{\beta }(B_{\varphi })B_{\varphi }}{g}
\end{equation}
while $B_{\varphi }$ is a solution of the equation
\begin{equation}
-\frac{\varepsilon ^{2}}{g}\varphi _{\beta }^{2}(B_{\varphi })B_{\varphi
}+\delta \varphi _{\alpha }(B_{\varphi })B_{\varphi }=0.  \label{eq}
\end{equation}
Over a wide range of parameters, this equation might possess five solutions
including the trivial one $B_{\varphi }=0$ . The other stationary points can
be found from the equation
\begin{equation}
\frac{\varepsilon ^{2}}{g\delta }(1+B_{\varphi }^{2})^{2}(1+k_{\alpha
}B_{\varphi }^{2})-(1+(k_{\beta }+1)B_{\varphi }^{2})^{2}=0.
\end{equation}
Since there are multiple stable solutions (see Fig.1 for $\varepsilon
=0.1,\;\delta =0.01,\;k_{\alpha }=k_{\beta }=1$), the system (\ref
{deterministic}) can exhibit hysteresis \cite{Tobias}.

\begin{center}
\includegraphics[width=0.5\textwidth]{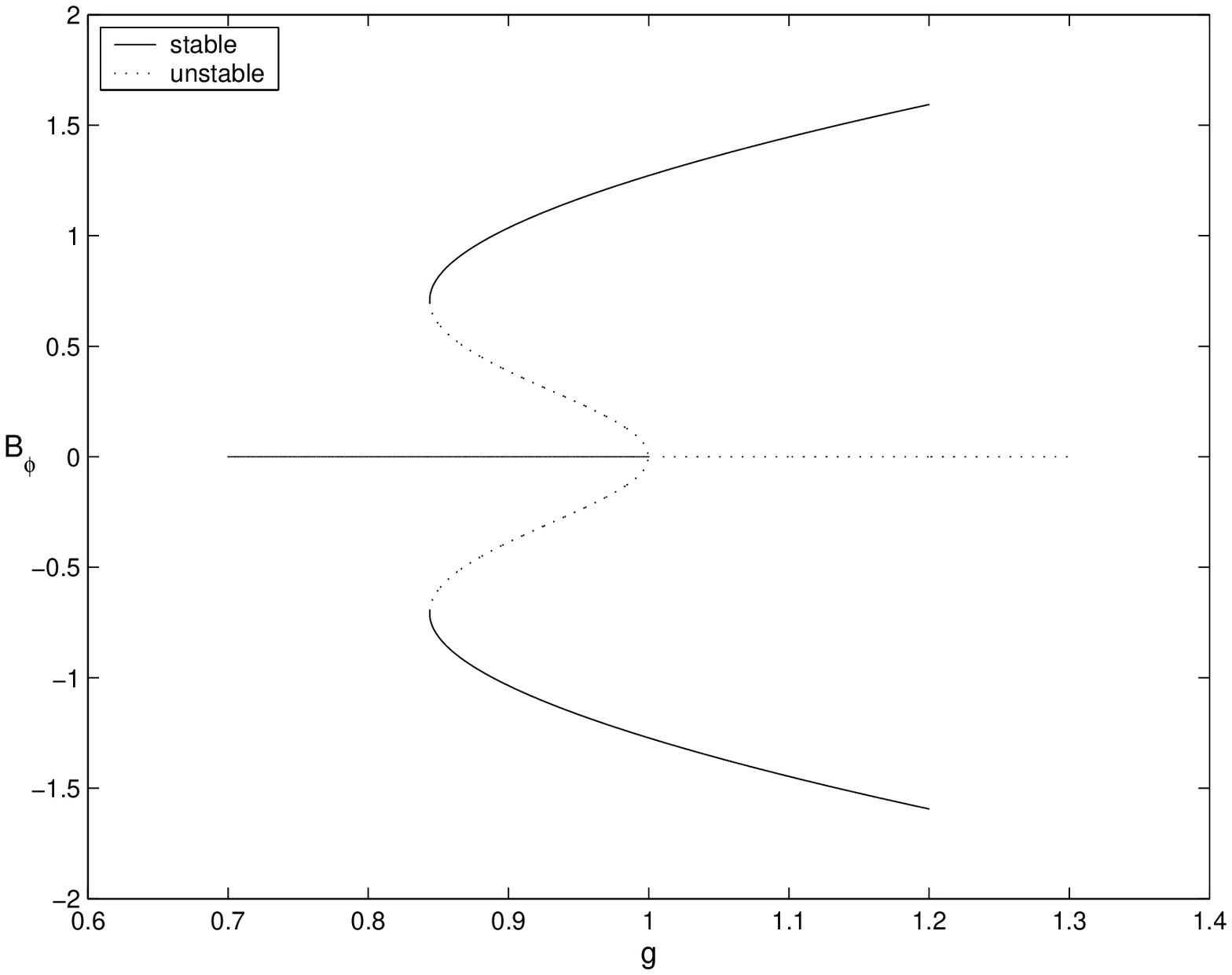}

Fig. 1. The dependence of the stationary points of the deterministic system (%
\ref{deterministic})\\[0pt]
on the parameter $g$ (subcritical bifurcation).
\end{center}

Of course, the stochastic system, in which such multiple stable states can
coexist (metastability), may exhibit random transitions between these
states. In general, the transition rate is inversely proportional to the
corresponding mean first passage time \cite{Gardiner,LH}. In the next
sections we discuss the peculiarities of such transitions with non-normal
effects.

The onset of the instability of the trivial equilibrium state $%
B_{r}=0,\;B_{\varphi }=0$ can be obtained from a standard linear stability
analysis. Linearization gives
\begin{equation}
\frac{d\mathbf{B}}{dt}=\mathbf{AB},  \label{onset}
\end{equation}
where
\begin{equation}
\mathbf{A}=\left[
\begin{array}{cc}
-\varepsilon & -\delta \\
-g & -\varepsilon
\end{array}
\right] ,\;\;\mathbf{B}=\left[
\begin{array}{c}
B_{r} \\
B_{\varphi }
\end{array}
\right] .
\end{equation}
The matrix $\mathbf{A}$ has two eigenvalues
\begin{equation}
\lambda _{1}=-\varepsilon +\sqrt{g\delta },\qquad \lambda _{2}=-\varepsilon -%
\sqrt{g\delta }.
\end{equation}
The \textit{supercritical} instability condition ($\lambda _{1}>0)$ can be
written as $g\delta >\varepsilon ^{2}$. The \textit{subcritical} case
corresponds to $g\delta <\varepsilon ^{2}$. The main purpose of this paper
is to study the stochastic amplification of the magnetic field $\mathbf{B}$
in the \textit{subcritical} case.

Since $\delta $ and $\varepsilon $ are small parameters and $g\sim 1$, the
matrix $\mathbf{A}$ is a highly non-normal one. Recall that the matrix $%
\mathbf{A}$ is normal, if $\mathbf{AA}^{T}=\mathbf{A}^{T}\mathbf{A},$ where $%
^{T}$ denotes the Hermitian transpose, otherwise it is non-normal. Even in
the\textit{\ subcritical case }when both eigenvalues $\lambda _{1,2}$ are
negative, $B_{\varphi }$ exhibits a large degree of transient growth before
the exponential decay. The azimuthal component $B_{\varphi }(t)$, as a
solution of the system (\ref{onset}) with the initial conditions
\begin{equation}
B_{r}(0)=-2c\sqrt{\delta /g},\;\;B_{\varphi }(0)=0,  \label{initial}
\end{equation}
has the form
\begin{equation}
B_{\varphi }(t)=c(e^{\lambda _{1}t}-e^{\lambda _{2}t}).
\end{equation}
Thus $B_{\varphi }(t)$ exhibits large transient growth over a timescale of
order $\frac{1}{\lambda _{2}-\lambda _{1}}\ln \frac{\lambda _{1}}{\lambda
_{2}}$ before decaying exponentially. In other words the transient growth
causes a temporary exit from the basin of attraction of the linearly stable
solution $(0,0)$. In Fig. 2 we plot phase portraits of the system (\ref
{deterministic}) for $\varepsilon =0.1,\;\delta =0.01,\;k_{\alpha }=k_{\beta
}=1$.

\begin{center}
\includegraphics[width=0.45\textwidth]{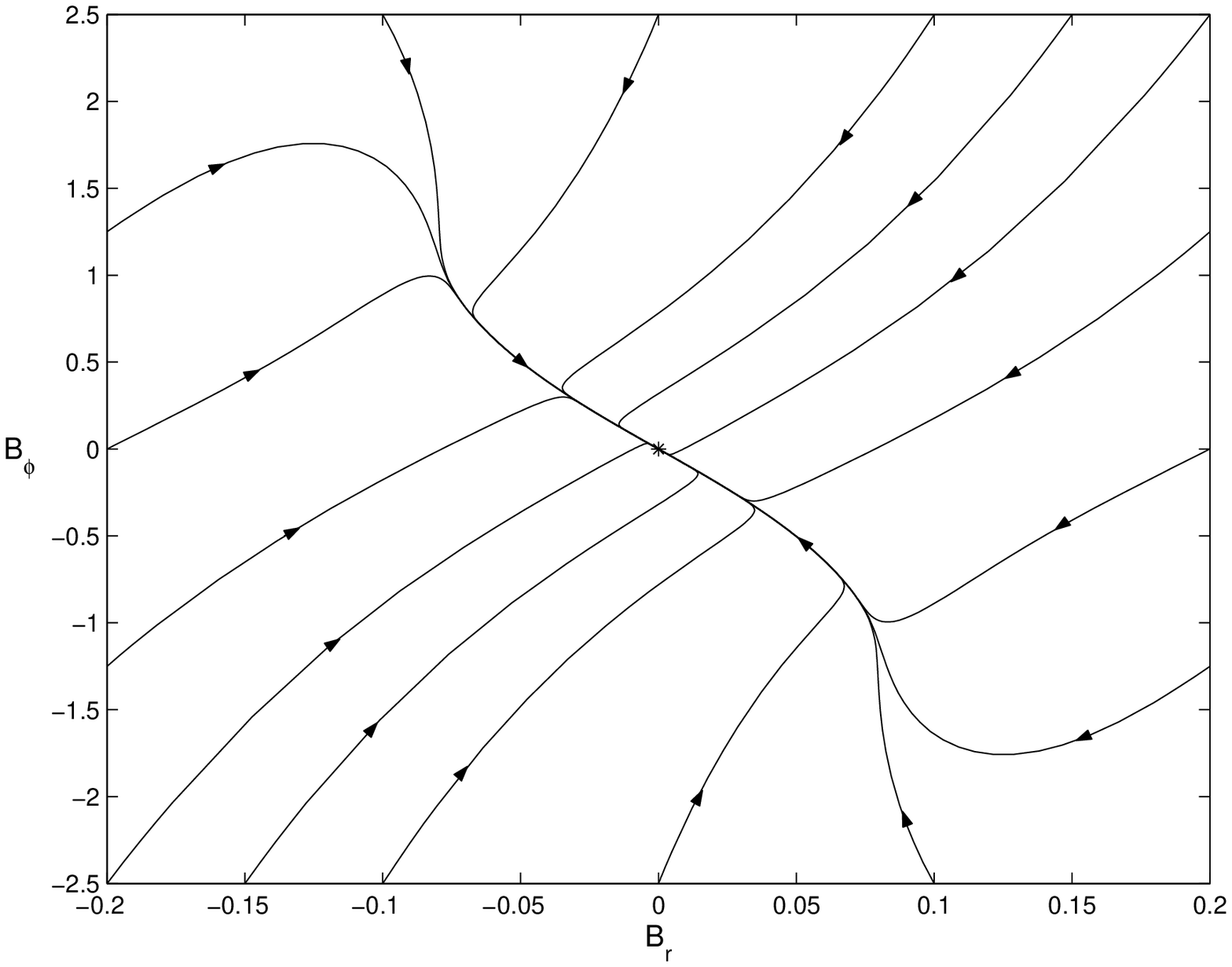} %
\includegraphics[width=0.45\textwidth]{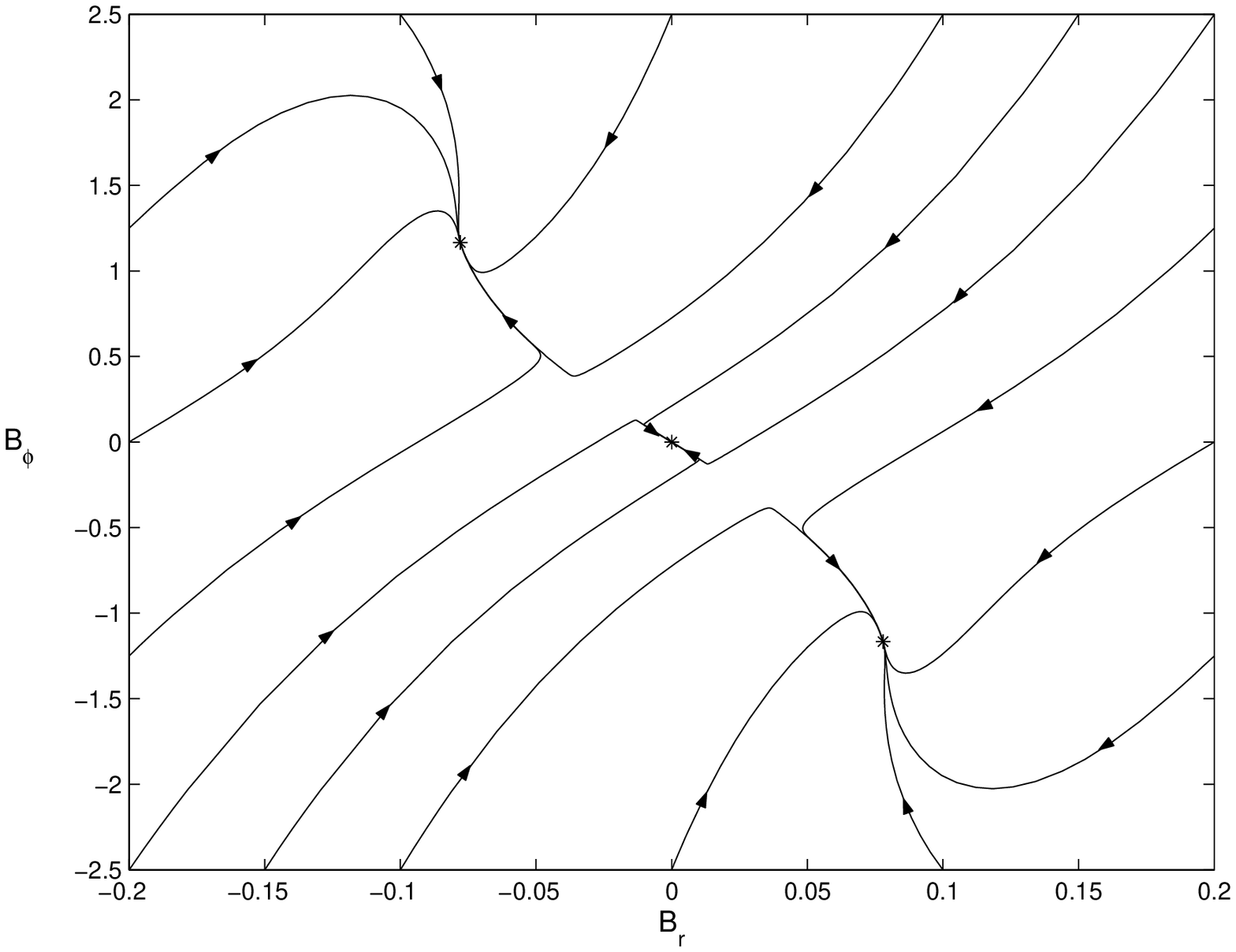} \vspace*{-3mm}

(a) $g=0.8$ one stable stationary point $(0,0)$ \hskip 1cm (b) $g=0.95$
three stable stationary points (subcritical case)

\vspace*{3mm} \includegraphics[width=0.45\textwidth]{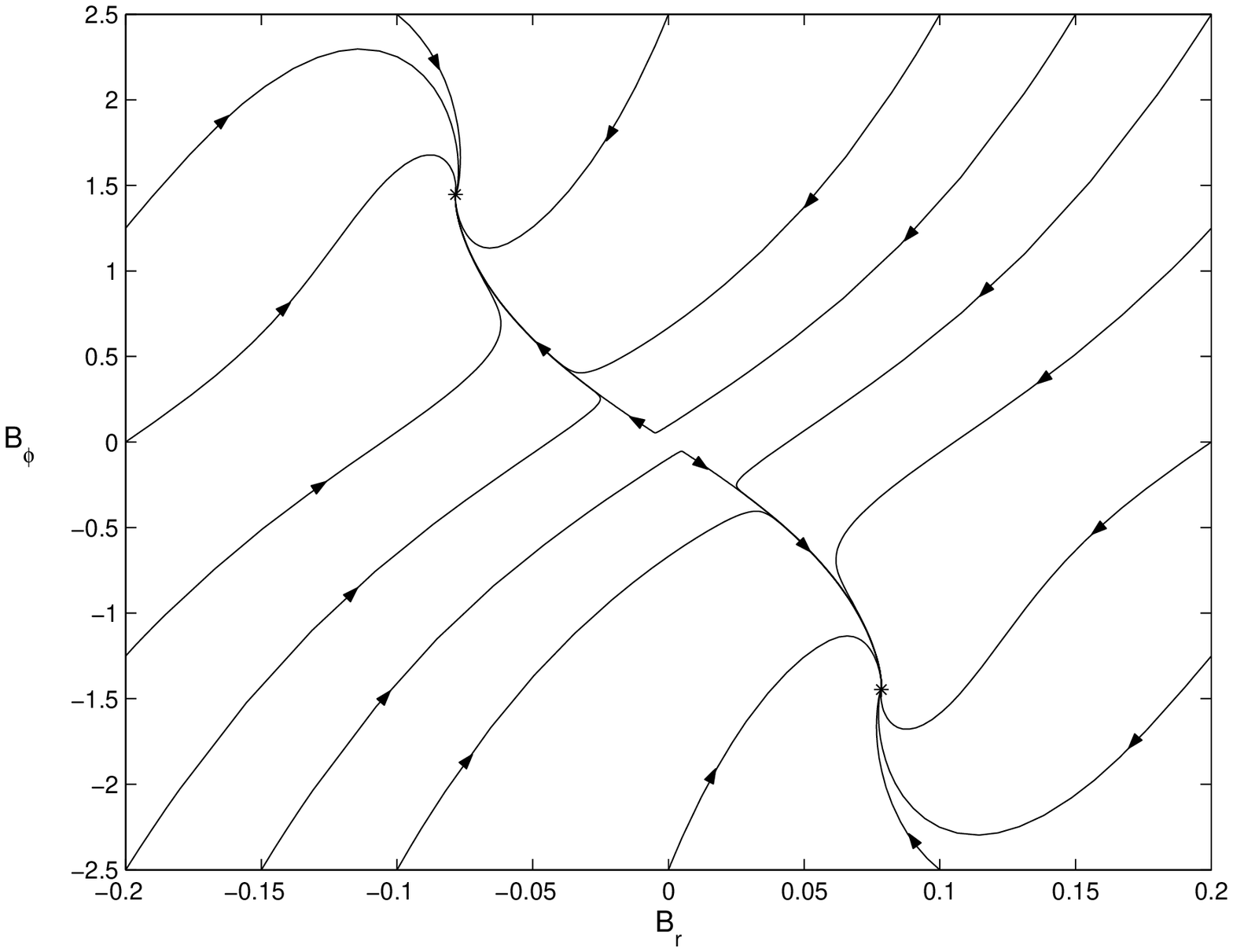}
\vspace*{-3mm}

\hskip 1cm (c) $g=1.1$ two stable stationary points (supercritical case)
\vspace*{1mm}

Fig. 2. Phase portraits of deterministic system (\ref{deterministic})\\[0pt]
\end{center}

Similar deterministic low-dimensional models like (\ref{deterministic}) with
transient growth have been proposed in \cite{Trefethen,GS,GG2,GG3} to
explain the \textit{subcritical} transition in the Navier-Stokes equations
\cite{SH}. It will be interesting to analyze the so-called self-killing and
self-creating dynamos by Fuchs, R\"{a}dler, and Rheinhard \cite{FRR} in
terms of non-linear subcritical instability \cite{F1}.

\section{Analysis of linear stochastic system: subcritical case}

In this section we discuss the circumstances under which the magnetic field
of initially zero amplitude can experience sustained growth even in the
subcritical case (kinematic regime). The linear approximation of the
stochastic system (\ref{basic1}),(\ref{basic}) near equilibrium point $%
B_{r}=0,\;B_{\varphi }=0$ can be written as
\begin{equation}
\frac{d\mathbf{B}}{dt}=\mathbf{AB}+\sqrt{2\sigma (\mathbf{B})}\mathbf{a}%
\frac{dW}{dt},  \label{stlinear}
\end{equation}
\begin{equation}
\mathbf{A}=\left[
\begin{array}{cc}
-\varepsilon & -\delta \\
-g & -\varepsilon
\end{array}
\right] ,\;\;\mathbf{B}=\left[
\begin{array}{c}
B_{r} \\
B_{\varphi }
\end{array}
\right] ,\;\;\;\;\mathbf{a}=\left[
\begin{array}{c}
1 \\
0
\end{array}
\right] ,  \label{w}
\end{equation}
\begin{equation}
\sigma (\mathbf{B})=\sigma _{1}B_{\varphi }^{2}+\sigma _{2}=\sigma _{1}(%
\mathbf{B,SB})+\sigma _{2},\;\mathbf{\;S}=\left[
\begin{array}{cc}
0 & 0 \\
0 & 1
\end{array}
\right] .
\end{equation}
Here $W$ is the standard Wiener process.

The important statistical characteristics of this system are the second
moments, since they represent the energy of the magnetic field (see, for
example \cite{F1,FI1}). The second moments matrix\textbf{\ }$\mathbf{M}=E(%
\mathbf{BB}^{\top })$ is governed by system
\begin{equation}
\frac{d\mathbf{M}}{dt}=\mathbf{AM}+\mathbf{MA}^{\top }+2(\sigma _{1}\mathrm{%
tr}(\mathbf{MS})+\sigma _{2})\mathbf{aa}^{\top }.  \label{matrix}
\end{equation}
There exists a variety of formal derivations of the equations (\ref{matrix}%
). One of them can be found in Appendix A. It is also convenient to
represent the matrix $\mathbf{M}$ in the form
\begin{equation}
\mathbf{M}=\left[
\begin{array}{cc}
m_{1} & m_{2} \\
m_{2} & m_{3}
\end{array}
\right] ,
\end{equation}
where $m_{1}(t)=E(B_{r}^{2}),$ $m_{2}(t)=E(B_{r}B_{\varphi }),$ $%
m_{3}(t)=E(B_{\varphi }^{2}),$ and $E(\cdot )$ denotes the expectation.

\subsection{Additive noise}

In this section we study how the additive noise can amplify the magnetic
energy
\begin{equation}
E_{B}(t)\approx m_{3}(t)=E(B_{\varphi }^{2})  \label{ener}
\end{equation}
during the kinematic stage. Without multiplicative noise $(\sigma _{1}=0)$
the system (\ref{stlinear}) takes the form
\begin{equation}
\frac{d\mathbf{B}}{dt}=\mathbf{AB}+\sqrt{2\sigma _{2}}\mathbf{a}\frac{dW}{dt}%
.  \label{onlyad}
\end{equation}
It follows from (\ref{matrix}) that in the \textit{subcritical} case the
second moments matrix $\mathbf{M}(t)$ converges to a stationary value as $%
t\rightarrow \infty $ \ such that
\begin{equation}
m_{1}=\frac{(2\varepsilon ^{2}-g\delta )\sigma _{2}}{2\varepsilon
(\varepsilon ^{2}-g\delta )},\;\;\;\;\;\quad m_{2}=\frac{g\sigma _{2}}{%
2(g\delta -\varepsilon ^{2})},\;\;\;\;\;\;\quad m_{3}=\frac{g^{2}\sigma _{2}%
}{2\varepsilon (\varepsilon ^{2}-g\delta )}.  \label{stat}
\end{equation}
This result implies that additive noise effects, which occur naturally in
the turbulent conducting fluids, together with differential rotation provide
mechanisms which generate the large scale magnetic field in the subcritical
case. Let us consider the dependence of the stationary value $%
m_{3}=E(B_{\varphi }^{2})$ on the differential rotation parameter $g$. When $%
g$ varies from $0$ to $g_{crit}=\varepsilon ^{2}/\delta $ the function $%
m_{3}(g)$ increases such that $m_{3}(g)\rightarrow \infty $ as $g\rightarrow
g_{crit}=\varepsilon ^{2}/\delta .$ Note that the sensitivity of normal
dynamical systems to additive noise near the bifurcation point is a
well-known result \cite{Gardiner}. Let us discuss now the differences
between a non-normal system like (\ref{onlyad}) and the normal one in the
presence of additive noise. To understand this difference consider along
with the non-normal system (\ref{onlyad}) the corresponding scalar normal
stochastic equation
\begin{equation}
\frac{dB}{dt}=\lambda _{1}B+\sqrt{2\sigma _{2}}\frac{dW}{dt},\qquad \lambda
_{1}=-\varepsilon +\sqrt{g\delta }.  \label{O-U}
\end{equation}
In the \textit{subcritical} case ($\lambda _{1}<0),$ the solution of (\ref
{O-U}) is nothing else but the classical Ornstein-Uhlenbeck process \cite
{Gardiner,LH}. It is easy to find the stationary value of its second moment $%
m=E(B^{2}):$
\begin{equation}
m=\frac{\sigma _{2}}{\varepsilon -\sqrt{g\delta }}.  \label{mom}
\end{equation}
To assess the significance of the effect of non-normality with respect to
the sensitivity to additive noise, let us introduce the new parameter $k=%
\frac{m_{3}}{m}$ that can be interpreted as a stochastic non-normality
coefficient. If we consider the neighborhood of the bifurcation point $%
\lambda _{1}=0,$ it gives us the measure of sensitivity of the non-normal
system to noise compared to the normal one. By using (\ref{stat}) and (\ref
{mom}), one can find
\begin{equation}
k=\frac{g^{2}}{2\varepsilon (\varepsilon +\sqrt{g\delta })}
\end{equation}
At the bifurcation point, $g=$ $g_{crit}$ the value of this parameter is
\begin{equation}
k=\frac{g_{crit}^{2}}{4\varepsilon ^{2}}.
\end{equation}
Since $g_{crit}\sim 1$ it follows that $k\sim \varepsilon ^{-2}>>1,$ which
shows how sensitive the non-normal system (\ref{onlyad}) is compared to the
equivalent normal system (\ref{O-U}). We note that the level of second
moments maintained in stochastic non-normal dynamical systems associated
with linearly stable shear flows has been discussed in \cite{F0}.

\subsection{Multiplicative noise: stochastic instability}

In this section we discuss how the average magnetic energy $E_{B}(t)\approx
m_{3}=E(B_{\varphi }^{2})$ is amplified by the random fluctuations of the $%
\alpha -$parameter in the \textit{subcritical }case (kinematic regime). In
particular we find the critical value of the multiplicative noise intensity $%
\sigma _{cr}$ such that for all values of $\sigma _{1}>\sigma _{cr}$ the
energy $E_{B}(t)$ grows as $\exp (\lambda t).$ during the early kinematic
stage.

Consider the system (\ref{stlinear}) with multiplicative noise only
\begin{equation}
\frac{d\mathbf{B}}{dt}=\mathbf{AB}+\sqrt{2\sigma _{1}(\mathbf{B,SB})}\mathbf{%
a}\frac{dW}{dt}.  \label{mult}
\end{equation}
It is well known that the second moments matrix of the system (\ref{stlinear}%
) converges to a stationary value as $t\rightarrow \infty $ if and only if
the equilibrium $\mathbf{B}=0$ of system (\ref{mult}) is exponentially
stable in the mean square sense (EMS-stable). EMS-stability means that \hskip%
1mm$E\Vert \mathbf{B}(t)\Vert ^{2}\leq K\exp {(-lt)}E\Vert \mathbf{B}%
(0)\Vert ,\;\;K,l>0$.

The equilibrium $\mathbf{B}=0$ of system (\ref{mult}) is EMS-stable if and
only if \newline
a) the equilibrium $\mathbf{B}=0$ of the deterministic system (\ref
{deterministic}) is asymptotically stable;\newline
b) $\mathrm{tr}(\mathbf{MS})<1,$ where $\mathbf{M}$ is the second moments
stationary matrix for the system
\begin{equation}
\frac{d\mathbf{B}}{dt}=\mathbf{AB}+\sqrt{2\sigma _{1}}\mathbf{a}\frac{dW}{dt}%
.  \label{lin_ad}
\end{equation}
One can find the proof of this result in Appendix B. This criterion (see
\cite{R}) reduces the linear stability analysis of a system with
multiplicative noise to that of the second moments stationary matrix of the
corresponding system with the additive noise only.

For the \textit{subcritical} case, it follows from this theorem that the
system (\ref{stlinear}) is EMS-stable if
\begin{equation}
g^{2}\sigma _{1}<2\varepsilon (\varepsilon ^{2}-g\delta ).
\end{equation}
The critical value of the multiplicative noise intensity $\sigma _{1}$ is
\begin{equation}
\sigma _{cr}=\frac{2\varepsilon (\varepsilon ^{2}-g\delta )}{g^{2}}.
\label{critical}
\end{equation}
If $\sigma _{1}>\sigma _{cr}$ then for any additive noise intensity $\sigma
_{2}>0$ the second moments of the system (\ref{mult}) tend to infinity as $%
t\rightarrow \infty $. This means that the random trajectories of the
nonlinear system (\ref{basic1}),(\ref{basic}) leave the basin of attraction
of zero equilibrium $B_{r}=0,\;B_{\varphi }=0$ . The understanding of the
stabilization of this growth should involve the numerical solution of the
full non-linear problem given by (\ref{basic1}),(\ref{basic}).

It should be noted that the idea of a magnetic generation by differential
rotation (the $\omega -$effect) is widely accepted, but the $\alpha -$effect
is still considered as controversial. It follows from the above analysis
that the amplification of magnetic energy might happen even in the case when
the mean value of $\alpha $ is zero. It should be noted that this result is
different from analogous in \cite{Sil} since we consider here the second
moments of the random magnetic field. Certainly, the predictions obtained
here for the simplified stochastic model should be ultimately extended to
more realistic systems of partial differential equations (\ref{main}).
However, for now we use the no$-z$ model as a simplified tool to observe the
noise effects on the generation of magnetic energy in the \textit{%
subcritical }case.

\section{Numerical analysis: nonlinear case}

So far we have concentrated on the linear stochastic instability where the
second moments grow exponentially without limit (kinematic regime).
Obviously, in the non-linear case, if we take into account the backreaction
which suppresses the effective dissipation $\beta ,$ and the $\alpha -$%
effect, one can expect an entirely different global behavior.

In this section, we perform simulations of random trajectories of the
nonlinear dynamical system (\ref{basic1}),(\ref{basic}) for $\varepsilon
=0.1,\;\delta =0.01,\;k_{\alpha }=k_{\beta }=1.$ In this case the critical
value of the differential rotation parameter $g_{crit}$ is $1.$ Since we are
interested in the \textit{subcritical }case, $g<1,$ we choose $g$ to be $%
0.99 $. It follows from (\ref{critical}) that for this set of parameters the
critical value of multiplicative noise intensity $\sigma _{cr} = 2\cdot
10^{-5}$. Let us emphasize again that the main reason why $\sigma _{cr}$ is
so small is because of the high level of non-normality of the dynamical
system (\ref{basic1}),(\ref{basic}).

Fig. 3 demonstrates qualitative changes in the shapes of the
probability density function (pdf) of $B_{\varphi }$  for fixed
$t$ as the intensity of
multiplicative noise $\sigma _{1}$ increases ($\sigma _{2}=0.5\cdot 10^{-7}$%
).

\begin{center}
\includegraphics[width=0.5\textwidth]{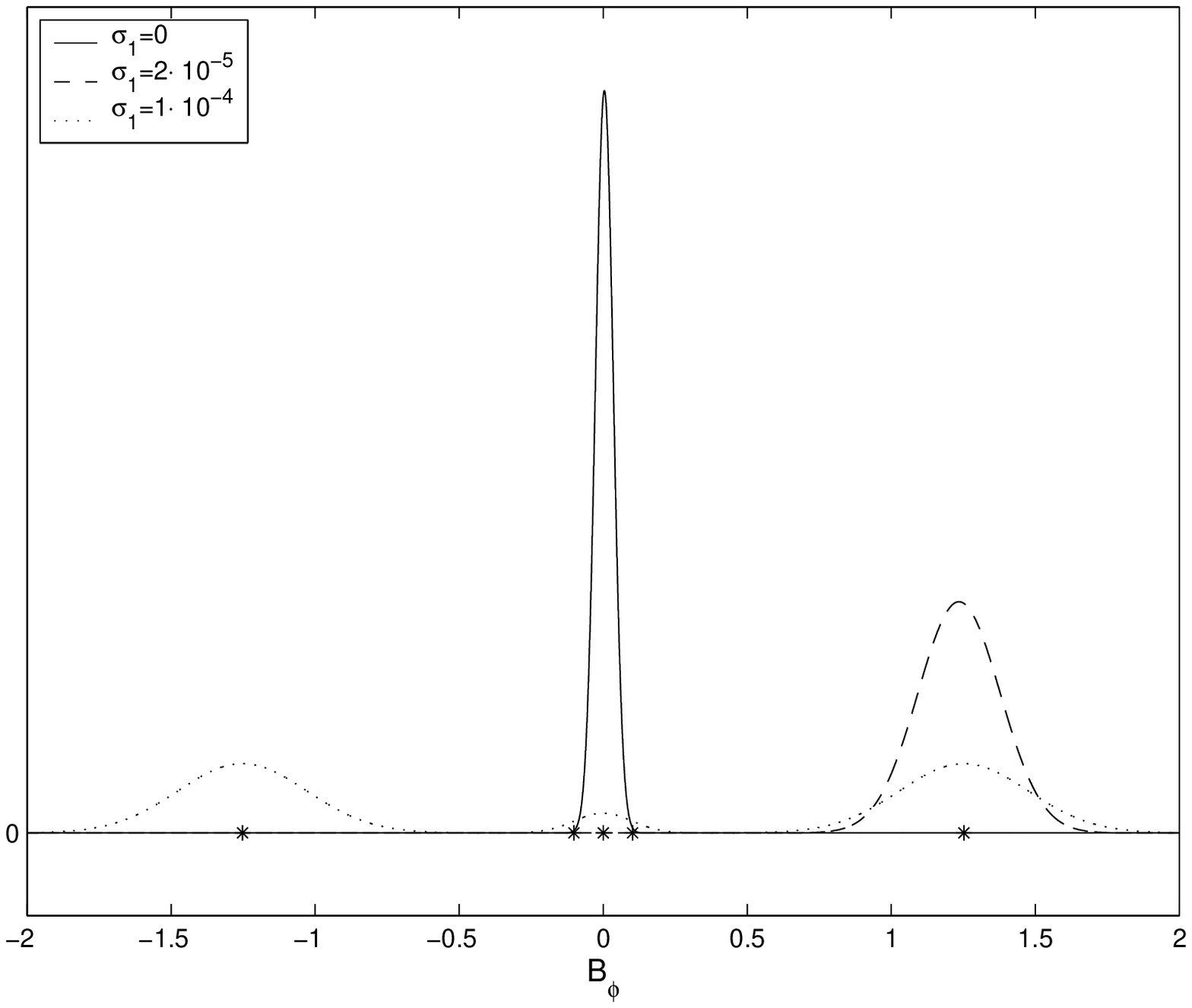}

Fig. 3. Probability density functions for different values of the
intensity of multiplicative noise.
\end{center}

Here asterisks mark the positions of the equilibrium points for
the deterministic system. One can see that there are three
qualitatively different regimes depending on the value $\sigma
_{1}.$ For very small values of $\sigma
_{1}$ the pdf is concentrated around the equilibrium point $%
B_{\varphi }=0.$ Stochastic trajectories of
(\ref{basic1}),(\ref{basic}) calculated by direct numerical
simulations are shown in Figs. 4 and 5.

\begin{center}
\includegraphics[width=0.6\textwidth]{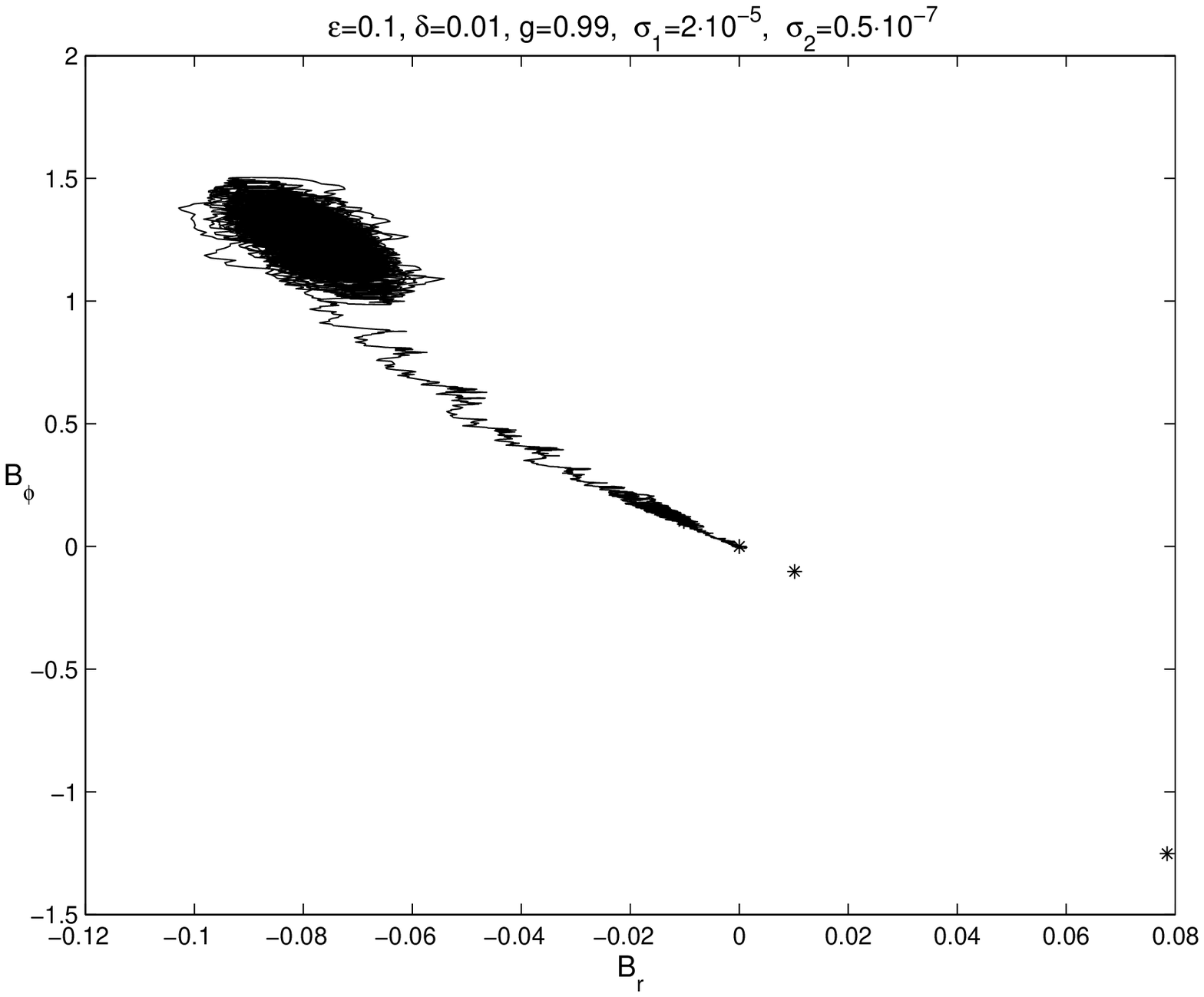}

Fig. 4. Stochastic trajectory escapes from zero equilibrium point and
concentrates near non-zero equilibrium.

\vspace*{1.0cm}

\includegraphics[width=0.6\textwidth]{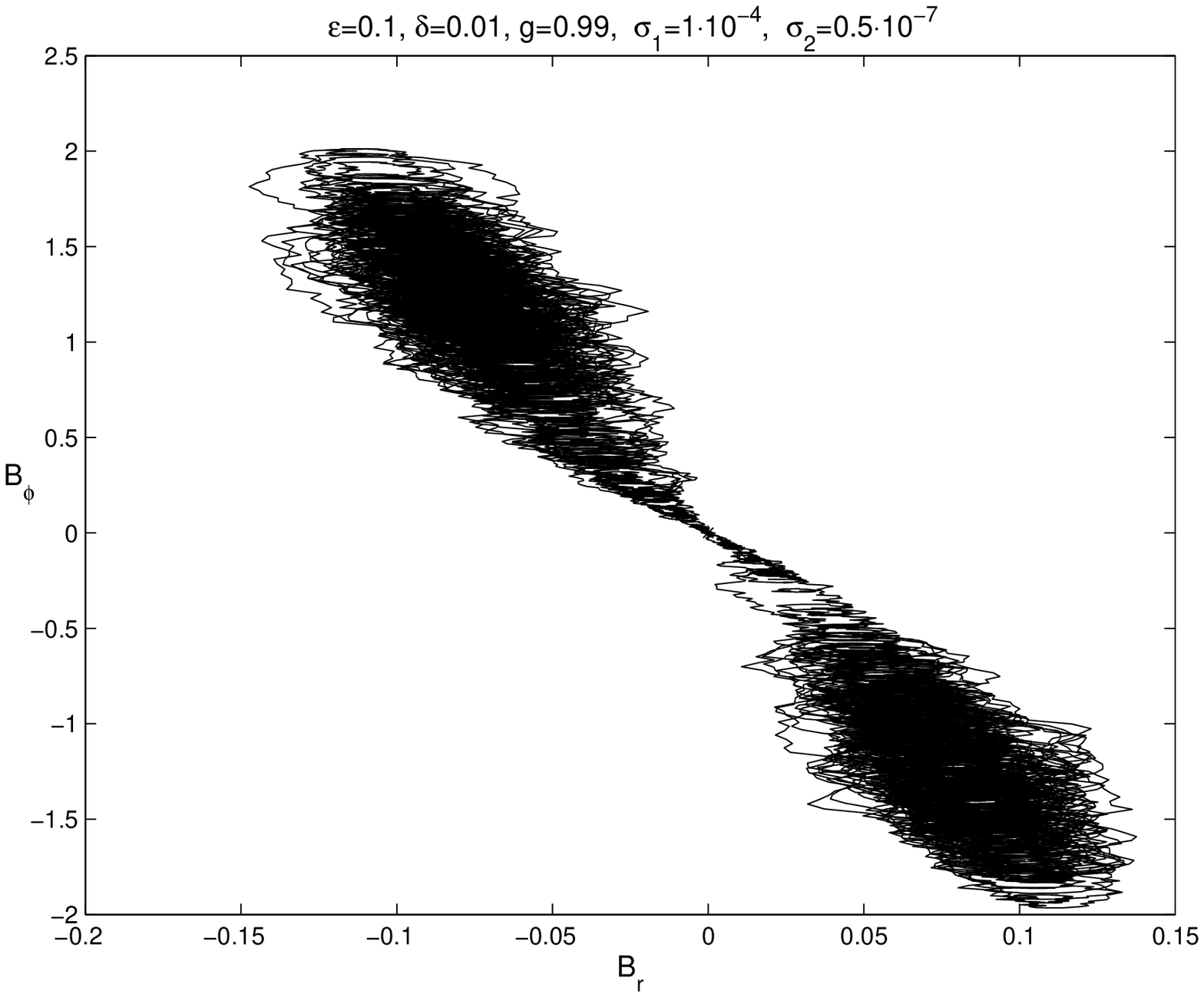}

Fig. 5. Stochastic trajectory moves around all stationary points.
\end{center}

One can see that as $\sigma _{1}$ increases, the trajectories (see Fig. 4)
escape from the domain of attraction of the zero equilibrium point and
concentrate in the vicinity of the non-zero stable equilibrium point.
Further increase of $\sigma _{1}$ leads to quite complicated dynamics when
the stochastic trajectory (see Fig. 5) moves around all deterministic
stationary points.

\section{Discussion and conclusions}

We have studied the stochastic amplification of large scale magnetic fields
in a differentially rotating system in a subcritical regime and discussed
the possible implications of this for the magnetic field in galaxies. The
main purpose was to address the stochastic generation that cannot be
explained by traditional linear eigenvalue analysis. We have chosen the
simplified stochastic $\alpha \Omega -$dynamo model for galaxies in a
thin-disk approximation and thereby concentrated on the influence of\
additive and multiplicative noises along with non-normality on the
amplification of the magnetic field in the subcritical case (when the dynamo
number is less than critical). In the linear case, we have derived the
equations for second moments describing the magnetic energy and demonstrated
the important differences between the non-normal system and the normal one
under the influence of additive noise. For the multiplicative noise, the
criteria for the stochastic instability during the early kinematic stage was
established in terms of the critical value of noise intensity due to $\alpha
$-fluctuations. In the non-linear case, we have performed numerical
simulations of non-linear stochastic differential equations for the $\alpha
\Omega -$dynamo and found a series of noise induced phase transitions:
qualitative changes in the behavior of the trajectories due to the increase
in the noise intensity parameter.

It should be noted that the equations (\ref{basic1}),(\ref{basic}) as
applied to galaxies are a theoretical simplification, but the do provide a
useful framework for understanding the effect of random fluctuations. Our
finding for the stochastic parametric instability can be straightforwardly
applied to partial differential equations (\ref{main}) in which case one has
to consider the coupled system for eigenmodes \cite{Hoyng}. In this respect
the two equations (\ref{basic1}),(\ref{basic}) can be regarded as a
dynamical system for first eigenmodes (order parameters), where the
influence of other degrees of freedom is parameterized by additive and
multiplicative noises. We believe that our theory can also be extended and
applied (after some modifications) to the solar dynamo in which a toroidal
field is generated by the action of a shear flow (typical non-normal effect).

\textbf{Acknowledgment}. In this project we benefited from the financial
support of this work by the Royal Society-Russia-UK Joint Project Grant. SF,
who was supported by Center of Turbulence Research, Stanford, is grateful to
Parviz Moin and Heinz Pitsch for a hospitality and fruitful discussions.

\renewcommand{\theequation}{A-\arabic{equation}} \setcounter{equation}{0}

\section*{Appendix A}

\textbf{Formal derivation of the equation for the second moments matrix}

Let ${\scriptstyle\Delta }\mathbf{B}=\mathbf{B}(t+{\scriptstyle\Delta }t)-%
\mathbf{B}(t)$. It follows formally from (\ref{stlinear}) that ${\scriptstyle%
\Delta }\mathbf{B}=\mathbf{AB}{\scriptstyle\Delta }t+{\mathbf{v}}{%
\scriptstyle\Delta }W,$ where $\mathbf{B}=\mathbf{B}(t),{\mathbf{v}}=\sqrt{%
2\sigma (\mathbf{B}(t))}\cdot \mathbf{a}$ or
\begin{equation}
\mathbf{B}(t+{\scriptstyle\Delta }t)=\mathbf{B}+\mathbf{AB}{\scriptstyle%
\Delta }t+\mathbf{v}{\scriptstyle\Delta }W  \label{A1}
\end{equation}
Equation (\ref{A1}) implies
\[
\mathbf{B}(t+{\scriptstyle\Delta }t)\mathbf{B}^{\top }(t+{\scriptstyle\Delta
}t)=(\mathbf{B}+\mathbf{AB}{\scriptstyle\Delta }t+{\mathbf{v}}{\scriptstyle%
\Delta }W)(\mathbf{B}^{\top }+\mathbf{B}^{\top }\mathbf{A}^{\top }{%
\scriptstyle\Delta }t+{\mathbf{v}}^{\top }{\scriptstyle\Delta }W)=
\]
\[
=\mathbf{BB}^{\top }+(\mathbf{ABB}^{\top }+\mathbf{BB}^{\top }\mathbf{A}%
^{\top }){\scriptstyle\Delta }t+\mathbf{vv}^{\top }({\scriptstyle\Delta }%
W)^{2}+(\mathbf{v}\mathbf{B}^{\top }+\mathbf{B}\mathbf{v}^{\top }){%
\scriptstyle\Delta }W+\newline
\]
\begin{equation}
+(\mathbf{AB}\mathbf{v}^{\top }+\mathbf{v}\mathbf{B}^{\top }\mathbf{A}^{\top
}){\scriptstyle\Delta }t{\scriptstyle\Delta }W+\mathbf{ABB}^{\top }\mathbf{A}%
^{\top }({\scriptstyle\Delta }t)^{2}  \label{A2}
\end{equation}
Taking into account the standard relations $E{\scriptstyle\Delta }W=0,E({%
\scriptstyle\Delta }W)^{2}={\scriptstyle\Delta }t$ for a Wiener process and $%
E(\mathbf{vv}^{\top })=2(\sigma _{1}tr(\mathbf{MS})+\sigma _{2})\mathbf{aa}%
^{\top }$ we can find from (\ref{A2}) for $\mathbf{M}(t)=E(\mathbf{B}(t)%
\mathbf{B}^{\top }(t))$ the following equation
\begin{equation}
{\scriptstyle\Delta }\mathbf{M}=\mathbf{M}(t+{\scriptstyle\Delta }t)-\mathbf{%
M}(t)=(\mathbf{AM}+\mathbf{MA}^{\top }+2(\sigma _{1}tr(\mathbf{MS})+\sigma
_{2})\mathbf{aa}^{\top }){\scriptstyle\Delta }t+\mathbf{AMA}^{\top }({%
\scriptstyle\Delta }t)^{2}  \label{A3}
\end{equation}
Now system (\ref{matrix}) follows from (\ref{A3}) immediately.

\renewcommand{\theequation}{B-\arabic{equation}} \setcounter{equation}{0}

\section*{Appendix B}

\textit{Necessity.} From EMS-stability of (\ref{mult}) it follows that there
exists a stationary second moments matrix $\mathbf{M}$ of system (\ref
{stlinear}) satisfying the equation
\begin{equation}
\mathbf{AM}+\mathbf{MA}^{\top }+2(\sigma _{1}\mathrm{tr}(\mathbf{MS})+\sigma
_{2})\mathbf{aa}^{\top }=0.  \label{A4}
\end{equation}
One can find for $\sigma _{1}>0$ that the matrix
\begin{equation}
\bar{\mathbf{M}}=\mathbf{M}/\left( \mathrm{tr}(\mathbf{MS})+\frac{\sigma _{2}%
}{\sigma _{1}}\right)
\end{equation}
is a solution of the equation
\begin{equation}
\mathbf{A}{\bar{\mathbf{M}}}+{\bar{\mathbf{M}}}\mathbf{A}^{\top }+2\sigma
_{1}\mathbf{aa}^{\top }=0.  \label{A5}
\end{equation}
Note that $\bar{\mathbf{M}}$ is a stationary second moments matrix for the
system (\ref{lin_ad}). The obvious inequality
\begin{equation}
\mathrm{tr}(\bar{\mathbf{M}}\mathbf{S})\;=\;\frac{\mathrm{tr}({\mathbf{M}}%
\mathbf{S})}{\mathrm{tr}({\mathbf{M}}\mathbf{S})+\frac{\sigma _{2}}{\sigma
_{1}}}<1
\end{equation}
proves necessity.

\textit{Sufficiency.} Consider a stationary second moments matrix $\mathbf{M}
$ of system (\ref{lin_ad}) satisfying equation (\ref{A5}) and the inequality
$\mathrm{tr}(\mathbf{MS})<1.$ It means for sufficiently small $\varepsilon
>0 $ a solution $\mathbf{M}_{\varepsilon }$ of the equation
\begin{equation}
\mathbf{AM}_{\varepsilon }+\mathbf{M}_{\varepsilon }\mathbf{A}^{\top
}+2\sigma _{1}\mathbf{aa}^{\top }+2\varepsilon \mathbf{I}=0,\;\;\mathbf{I}=%
\left[
\begin{array}{cc}
1 & 0 \\
0 & 1
\end{array}
\right]  \label{A6}
\end{equation}
satisfies the inequality
\begin{equation}
\mathrm{tr}(\mathbf{M}_{\varepsilon }S)<1  \label{A7}
\end{equation}
too. From (\ref{A6}), (\ref{A7}) it follows that
\begin{equation}
\mathbf{AM}_{\varepsilon }+\mathbf{M}_{\varepsilon }\mathbf{A}^{\top
}+2(\sigma _{1}\mathrm{tr}(\mathbf{M}_{\varepsilon }S)+\bar{\sigma}_{2})%
\mathbf{aa}^{\top }+2\varepsilon \mathbf{I}=0  \label{A8}
\end{equation}
where ${\bar{\sigma}}_{2}=\sigma _{1}(1-\mathrm{tr}(\mathbf{M}_{\varepsilon }%
\mathbf{S}))>0.$ Formula (\ref{A8}) shows that $\mathbf{M}_{\varepsilon }$
is a stationary second moments matrix for the stochastic system
\begin{equation}
\frac{d\mathbf{B}}{dt}=\mathbf{AB}+\sqrt{2(\sigma _{1}(\mathbf{B},\mathbf{SB}%
)+\bar{\sigma}_{2})}\mathbf{a}\frac{dW}{dt}+\sqrt{2\varepsilon }\frac{d\xi }{%
dt}  \label{A9}
\end{equation}
where $\xi $ is the independent two-dimensional standard Wiener process. The
existence of a stationary second moments matrix for the stochastic system (%
\ref{A9}) with two-dimensional nondegenerate additive noise of intensity $%
\varepsilon >0$ proves EMS-stability of the system (\ref{mult}). The details
of the proof can be found in \cite{R}.

\newpage

\end{document}